\documentclass[english,12pt,aps,prd,a4paper,preprintnumbers,floatfix,nofootinbib,showpacs,superscriptaddress, notitlepage]{revtex4-1} 
 \pdfoutput=1
\usepackage[usenames,dvipsnames]{color}  
\usepackage{graphicx}
\usepackage{pifont}
\usepackage{caption}
\usepackage{subcaption}
\usepackage{amsmath}
\usepackage{amssymb}
\usepackage[colorlinks=true,citecolor=darkred,urlcolor=darkred, pdfborder={0 0 0}]{hyperref}
\usepackage[normalem]{ulem}

\usepackage[T1]{fontenc}
\usepackage[latin9]{inputenc}
\usepackage{array}
\usepackage{booktabs}
\usepackage{mathrsfs}
\usepackage{multirow}
\usepackage{tabularx}
\usepackage[export]{adjustbox}
\usepackage{float}
\usepackage{multirow}

\definecolor{darkred}{rgb}{0.6,0,0}
\definecolor{dgreen}{rgb}{0,0.5,0}

\definecolor{linkcolor}{rgb}{0,0,0.5}

\def\gsim{\raise0.3ex\hbox{$\;>$\kern-0.75em\raise-1.1ex\hbox{$\sim\;$}}}
\def\lsim{\raise0.3ex\hbox{$\;<$\kern-0.75em\raise-1.1ex\hbox{$\sim\;$}}}

\def\beqn#1{\begin{equation}\label{#1}}
\def\eeqn{\end{equation}}

\def\beqa#1{\begin{eqnarray}\label{#1}}
\def\eeqa{\end{eqnarray}}

\def\Z2{$\mathcal{Z_2}$}

\newcommand {\ignore}[1]{}

\def\cevns{CE$\nu$NS~}
\def\321{$\mathrm{SU(3) \otimes SU(2) \otimes U(1)}$ }

\newcommand{\AddrAHEP}{%
  AHEP Group, Institut de F\'{i}sica Corpuscular --
  CSIC/Universitat de Val\`{e}ncia \\ 
 C/ Catedr\'atico Jos\'e Beltr\'an, 2 E-46980 Paterna, Spain}

\newcommand{\AddrFisteo}{%
Departament de F\'{i}sica Te\'orica, Universitat de Val\`encia, Burjassot 46100, Spain}

\newcommand{\AddrMiranda}{%
Departamento de F\'{\i}sica, Centro de Investigaci\'on
  y de Estudios Avanzados del IPN,\\ Apartado Postal 14-740 07000 Mexico,
  Distrito Federal, Mexico}

\newcommand{\AddrIoannina}{%
Division of Theoretical Physics, University of  Ioannina, GR 45110 Ioannina, Greece}

\begin{document}

\bibliographystyle{unsrt}   

\title{\boldmath \color{BrickRed} Future CEvNS experiments as probes of \\
  lepton unitarity and light sterile neutrinos} 

\author{O. G. Miranda}\email{omr@fis.cinvestav.mx}\affiliation{\AddrMiranda}

\author{D.K. Papoulias}\email{d.papoulias@uoi.gr}\affiliation{\AddrIoannina}

\author{O. Sanders}\email{osanders@fis.cinvestav.mx}
\affiliation{\AddrMiranda}

\author{M. T\'ortola}\email{mariam@ific.uv.es}
\affiliation{\AddrFisteo}\affiliation{\AddrAHEP}

\author{J. W. F. Valle}\email{valle@ific.uv.es}
\affiliation{\AddrAHEP}

\begin{abstract}
\vskip .5cm
We determine the sensitivities of short-baseline coherent elastic neutrino-nucleus scattering (CE$\nu$NS) experiments using a pion decay at rest neutrino source as a probe for nonunitarity in the lepton sector, as expected in low-scale type-I seesaw schemes.
We also identify the best configuration for probing light sterile neutrinos at future ton-scale liquid argon \cevns experiments, estimating the projected sensitivities on the sterile neutrino parameters. 
Possible experimental setups at the Spallation Neutron Source, Lujan facility and the European Spallation Source are discussed.
Provided that systematic uncertainties remain under control, we find that \cevns experiments will be competitive with oscillation measurements in the long run.

\end{abstract}

\maketitle

\section{Introduction}
\label{sec:intro}

The three-neutrino paradigm has been put on rather solid ground from the interpretation of solar and atmospheric oscillation data and the
complementary results from reactor and accelerator neutrino studies~\cite{deSalas:2017kay}.
Underpinning the precise way by which neutrinos get mass is one of the main current challenges in particle physics.
One of the leading ideas is that neutrino mass generation proceeds through the mediation of new heavy fermion states, such as in variants of the so-called type-I seesaw mechanism.
Since they carry no anomaly, isosinglet ``right-handed'' mediators can come in an arbitrary  number in the Standard Model (SM), so one can envisage low-scale seesaw realizations, where the mediators can lie at the TeV scale with potentially sizable mixing with the light neutrinos~\cite{Mohapatra:1986bd,Akhmedov:1995vm,Akhmedov:1995ip,Malinsky:2005bi}.
The admixture of heavy lepton messengers implies that the charged-current weak interaction mixing matrix has a rectangular form~\cite{Schechter:1980gr},
leading to unitarity violation, as these heavy states are not kinematically accessible. Likewise, one expects universality violation effects.
The associated processes could take place below~\cite{Gronau:1984ct},
  at~\cite{Bernabeu:1987gr,Rius:1989gk,Dittmar:1989yg}
    or above~\cite{Abreu:1996pa,Achard:2001qv,Aad:2015xaa,Sirunyan:2018mtv} the $Z$ boson mass scale.
In the context of neutrino propagation, the admixture of heavy neutrinos would clearly also imply deviations from unitarity, as the heavy states cannot take part in oscillations.

Unitarity violation in neutrino oscillations has been explicitly considered in Refs.~\cite{Valle:1987gv,Nunokawa:1996tg,Antusch:2006vwa,Forero:2011pc,Miranda:2016ptb,Blennow:2016jkn,Escrihuela:2015wra,Escrihuela:2016ube,Tang:2017khg}. 
It has been noticed that the extra $CP$ violation expected in these schemes can fake the one present within the simplest three-neutrino paradigm~\cite{Miranda:2016wdr}.
As a result, unitarity violation degrades the CP violation sensitivity expected at DUNE~\cite{Escrihuela:2016ube}.
Here we note that the subleading effects of such TeV-scale heavy neutrino mediators can also be probed in future liquid argon coherent elastic neutrino-nucleus scattering (CE$\nu$NS) experiments using muon decays as the neutrino source.

On the other hand, controversial anomalies such as those coming from recent reactor data, as well as those hinted by the LSND~\cite{Aguilar:2001ty} and MiniBooNE~\cite{Aguilar-Arevalo:2018gpe} experiments, inspired many phenomenological studies beyond the simplest three-neutrino oscillation picture~\cite{Abazajian:2012ys,Gariazzo:2017fdh,Dentler:2018sju}. These are based on the existence of a fourth \textit{light} sterile neutrino state, with  eV-scale mass ($m_{1,2,3} \ll m_4$).
Indeed, under certain circumstances, such as special symmetries~\cite{Peltoniemi:1992ss,Peltoniemi:1993ec}, one may expect such extra light sterile neutrinos to emerge in fermion mediator models of neutrino mass generation.

The importance of neutral-currents in oscillation physics has been known for a long time; see Refs.~\cite{Freedman:1973yd,Schechter:1980gr}.
The discovery of \cevns has now brought neutral-current-based experiments to center stage, as a competitive and complementary tool to shed light on fundamental neutrino parameters.
Facilities looking for \cevns have been recognized to be important probes of sterile neutrino oscillations for aboout a decade~\cite{Formaggio:2011jt,Anderson:2012pn,Kosmas:2017zbh}.
In 2017, the COHERENT Collaboration reported the first observation of \cevns on CsI~\cite{Akimov:2017ade} at the Spallation Neutron Source (SNS), a result recently confirmed by the same collaboration on a liquid argon detector~\cite{Akimov:2020pdx}. 
This prompted a new era with a wide range of physics applications concerning open questions within Refs.~\cite{Cadeddu:2017etk,Cadeddu:2018izq,Canas:2018rng,Cadeddu:2017etk,Papoulias:2019lfi, AristizabalSierra:2019zmy, Coloma:2020nhf, 1805471} and beyond the SM \cite{Liao:2017uzy,Dent:2017mpr,AristizabalSierra:2018eqm, Denton:2018xmq, Kosmas:2017tsq,Cadeddu:2018dux,Miranda:2019wdy,Parada:2019gvy, Dutta:2019eml,Coloma:2017ncl,Gonzalez-Garcia:2018dep,Canas:2019fjw, Papoulias:2019xaw, Coloma:2019mbs, Dutta:2020che, Flores:2020lji, Cadeddu:2020lky, Miranda:2020tif,Farzan:2018gtr,Abdullah:2018ykz,Brdar:2018qqj, Billard:2018jnl, AristizabalSierra:2019ufd, AristizabalSierra:2019ykk,  Miranda:2020zji}, including also dedicated sterile neutrino searches~\cite{Canas:2017umu,Blanco:2019vyp,Miranda:2019skf,Berryman:2019nvr}. 
 The field is thriving rapidly, with several experiments aiming to measure \cevns now in preparation worldwide (for a review see
Ref.~\cite{Akimov:2019wtg}), many of which are planning to employ large liquid argon detectors.  Here we quantify the prospects for
probing the effects of both light sterile neutrinos and heavy neutrino-mass mediators within future proposals employing large liquid
argon scintillation detectors. In particular, we concentrate on the next-generation detector subsystem of COHERENT, namely
CENNS~\cite{Akimov:2019xdj}, as well as on the Coherent Captain-Mills (CCM) experiment~\cite{CCM} at the Los Alamos Neutron Science Center
- Lujan facility, and on the \cevns program developed at the European Spallation Source (ESS)~\cite{Baxter:2019mcx}.

Our work is organized as follows. In Sec.~\ref{sec:theory} we present the required formalism for simulating \cevns signals and discuss the experimental sites considered.
In Sec.~\ref{sec:heavy-singl-neutr} we present our results concerning nonunitarity effects induced by new heavy neutrino admixtures and in Sec.~\ref{sec:light-ster-neutr} we discuss the sensitivities we have obtained on light sterile neutrinos. Finally,  we summarize and conclude in Sec.~\ref{sec:conclusions}.

\section{Simulating coherent elastic neutrino-nucleus scattering}
\label{sec:theory}

Our present research on indirect effects of heavy neutrino states or light sterile neutrinos is motivated by future neutral-current \cevns
measurement proposals.
Previous work on sterile neutrino constraints from
  the CsI COHERENT measurement can be found in Ref.~\cite{Kosmas:2017tsq}.
  We consider the process
$\nu_\alpha + (A,Z) \to \nu_\beta + (A,Z)$ where $A$ and $Z$ stand for
the mass and atomic number of a nucleus, respectively, while $E_\nu$ is
the neutrino energy and $\alpha, \beta$ represent the flavor index
($\alpha, \beta =e, \mu, \tau$). In this section, we summarize the
relevant formalism for simulating the expected \cevns signal and 
discuss the various experimental configurations considered in our
 analysis.

\subsection{Coherent elastic neutrino-nucleus scattering}

The \cevns cross section scales as $N^2$, where $N=A-Z$ is the number of neutrons and,
therefore, leads to an enhanced neutrino interaction cross section~\cite{Freedman:1973yd}.
The relevant \cevns experiments are mainly sensitive to the tiny recoils generated in a
scattering event. The differential cross section in terms of the nuclear
  recoil energy, $T_A$, is~\cite{Barranco:2005yy} 
\begin{equation}
\left(\frac{\mathrm{d} \sigma}{\mathrm{d} T_A}\right)_{\text{SM}} = \frac{G_F^2 m_A}{2 \pi}  (\mathcal{Q}_W)^2 \left[ 2 - \frac{2 T_A}{E_\nu} - \frac{m_A T_A}{E_\nu^2}\right]  \, .
\label{eq:xsec-cevns}
\end{equation}
Here, $G_F$ denotes the Fermi constant, $m_A$ is the nuclear mass, and $\mathcal{Q}^V_W$ is the  weak charge~\cite{Papoulias:2015vxa}
\begin{equation}
\mathcal{Q}_W =  \left[ \left(\frac12-2 \sin^2 \theta_W \right) Z F_{p}(Q^{2})  -\frac12 N F_{n}(Q^{2}) \right]  \, ,
\label{eq:Qw}
\end{equation}
written in terms of the weak mixing angle $\sin^2 \theta_W  = 0.2312$, taken in the $\overline{\text{MS}}$ scheme.
A coherence loss, due to the finite nuclear size, is incorporated
through the nuclear form factors for protons and neutrons,
$F_{p,n}(Q)^{2}$. Amongst the various available parametrizations in the
literature (for a summary see Ref.~\cite{Papoulias:2019xaw}), here we
employ the well-known Helm form factor, given by
\begin{equation}
  F_{p,n}(Q^2) = 3\frac{j_1(QR_0)}{QR_0}\exp(-Q^2s^2/2) ,
  \end{equation}
where the magnitude of the three-momentum transfer is
$Q=\sqrt{2m_AT_A}$, the spherical Bessel function of order one is
$j_1(x)=\sin(x)/x^2 - \cos(x)/x$, and
$R_0^2=\frac53(R_{p,n}^2-3s^2)$. For the relevant liquid argon detectors, the neutron and proton
rms radii take the values $R_n=3.36$~fm and $R_p=3.14$~fm, while the surface thickness is $s=0.9$~fm.

As for the incoming neutrino flux, at spallation source facilities a large number of protons is scattered
on a nuclear target (mercury for the SNS and tungsten for CCM and ESS), producing pions. The latter
propagate and subsequently decay at rest generating neutrinos ($\pi$-DAR neutrinos).
A monochromatic neutrino beam is produced from
$\pi^+ \to \mu^+ \nu_\mu $ (prompt flux, with lifetime $\tau=26$~ns) with a spectrum given by  
\begin{equation}
 \frac{\mathrm{d}\phi_{\nu _{\mu }} }{\mathrm{d}E_\nu }= \delta\left ( E_\nu-\frac{m_{\pi }^{2}-m_{\mu }^{2}}{2m_{\pi }} \right ) \, .
\label{eq:PromptFlux}
\end{equation}
The subsequent muon decay $\mu^+ \to \bar{\nu}_\mu e^+ \nu_e$ (delayed flux, $\tau=2.2~\mathrm{\mu s}$)  generates a beam composed of muon antineutrinos
\begin{equation}
\frac{\mathrm{d}\phi_{\overline{\nu} _{\mu }} }{\mathrm{d}E_\nu }=  \frac{64E_\nu^{2}}{m_{\mu }^{3}}\left ( \frac{3}{4}-\frac{E_\nu}{m_{\mu }} \right ) \, ,
\label{eq:DelFluxMuon}
\end{equation}
and electron neutrinos
\begin{equation}
\frac{\mathrm{d}\phi_{\nu _{e }} }{\mathrm{d}E_\nu }=  \frac{192E_\nu^{2}}{m_{\mu }^{3}}\left ( \frac{1}{2}-\frac{E_\nu}{m_{\mu }} \right ) \, . 
\label{eq:DelFluxEl}
\end{equation}

\subsection{Experimental sites} 
\label{sec: experiments}

Our present analysis will be focused on three prominent experiments
aiming to deploy large liquid argon detectors (see Table~\ref{tab:exps}) to measure
a \cevns signal at a $\pi$-DAR source. 
We first consider the next-generation CENNS detector of the COHERENT experiment at the
SNS~\cite{Akimov:2019xdj}, which is expected to replace the CENNS-10 detector that provided
the first detection of \cevns on argon~\cite{Akimov:2020pdx}.
The planned configuration will contain a 750~kg (610~kg fiducial) liquid argon scintillation detector
and will operate with a 20~keV threshold and a baseline of 28.4~m.
Another interesting experimental site is the proposed CCM experiment, located at Los Alamos National
Laboratory, in the Lujan facility. The CCM experiment plans to install a large 7 ton liquid argon detector
and is expected to achieve a 1 keV threshold~\cite{CCM}.
The detector will be placed 20~m from the source with the goal to search for sterile neutrinos.
Another promising facility is the ESS located in Lund, Sweden, that combines the world's most powerful
superconducting proton linac with an advanced hydrogen moderator, generating the most intense neutron
beam for different purposes.
Following the proposal in Ref.~\cite{Baxter:2019mcx}, we will assume two different configurations:
i) a first phase configuration with a 10~kg liquid argon detector and an ultra-low 0.1~keV threshold and
ii) a next-generation configuration with a 1 ton liquid argon detector and a
20~keV threshold, both located 20~m from the source.

The main difference among the ESS, SNS and Lujan facilities is that the former one is scheduled to reach a power of 5~MW with a goal energy of 2~GeV by 2023, while SNS (Lujan) will have a power of
1.3~MW~\footnote{A possible increase to 2.4~MW is feasible with a
  second target station at the SNS~\cite{Akimov:2019xdj}.} (80~kW).
This will lead to an about 1 order of magnitude increase in the ESS neutrino flux with respect to SNS,
resulting in a significantly faster accumulation of \cevns signal statistics in comparison with the other two facilities.
A second difference is the proton beam pulse timing: SNS provides 60 Hz of 1~$\mu$s-wide proton on target (POT) spills
while ESS can only offer 14 Hz of 2.8 ms spills, reducing the relative capability of separating
the neutrino flavors with timing information.
Finally, while the power of the proton beam at Lujan is 1--2 orders of magnitude smaller than in SNS
and ESS, it is worth mentioning that, in contrast to SNS, the CCM experiment can deploy very large
ton-scale detectors. This feature, together with the fact that the Lujan Center can achieve a shorter
beam time interval, makes the CCM experiment clearly complementary to the \cevns searches at SNS and ESS.

\begin{table}[t]
\begin{tabular}{|l|c|c|c|c|}
\hline
                                            & CENNS~\cite{Akimov:2019xdj} & CCM~\cite{CCM}   & ESS~\cite{Baxter:2019mcx}            \\
                                            \hline
mass                                        & 610 kg      & 7 ton & 10 kg (1 ton)  \\
threshold ($\mathrm{keV_{nr}}$)             & 20           & 1     & 0.1 (20)       \\
$N_\mathrm{POT}$ ($10^{23}$/yr)         & 1.5          & 0.177     & 2.8  \\
$r$                                         & 0.08         & 0.0425    & 0.3 \\
baseline (m)                                & 28.4         & 20    & 20       \\
\hline      
\end{tabular}
\caption{Experimental configurations assumed in the present work.}
\label{tab:exps}
\end{table}

\subsection{Statistical analysis}
Our statistical analysis is based on the expected number of events, simulated for each experiment.
For the case of CE$\nu$NS, the differential number of events is given by
\begin{equation}
\frac{\mathrm{d}N_x}{\mathrm{d}T_A} = \eta \, t_\text{run} N_\text{target} \sum_{\nu_\alpha} \int_{\sqrt{\frac{m_A T_A}{2}}}^{E_\nu^\mathrm{max}} \, \mathcal{A}(T_A) \, \frac{\mathrm{d}\phi_{\nu_\alpha}}{\mathrm{d} E_\nu} \left(\frac{\mathrm{d} \sigma}{\mathrm{d} T_A}\right)_{x} \, \mathrm{d}E_\nu \, ,
\label{eq:dNdTA}
\end{equation}
where $t_\text{run}$ is the data-taking time (we will assume  $t_\text{run}$ = 1 year),
$N_\text{target}$ is the number of nuclear targets in the detector,
and $x=\text{(SM, new)}$ denotes the type of interaction.
Here, $\eta$ denotes a normalization factor given by $\eta = rN_{\text{POT}}/4\pi L^{2}$, where $L$ is the
baseline, $N_{\text{POT}}$ is the number of delivered POT and $r$ is the number of
produced neutrinos per POT. Finally, the detector efficiency $\mathcal{A}(T_A)$ for the case of CENNS-610 is taken from Ref.~\cite{Akimov:2020czh}, while for ESS and CCM the efficiency is assumed to be unity (zero)   above (below) the corresponding recoil threshold shown in Table~\ref{tab:exps}.

Our statistical analysis will be mainly based on the $\chi^2$ function
\begin{equation}
\chi^2(\alpha_{11},|\alpha_{21}|,\alpha_{22}) = \underset{\mathtt{a}}{\mathrm{min}} \Bigg [\sum_i \frac{\left(N^i_{\text{SM}} - N^i_{\text{new}} [1+\mathtt{a}]\right)^2} {(\sigma^i_\text{stat})^2}   + \left( \frac{\mathtt{a}}{\sigma_{\text{sys}}}\right)^2 \Bigg] \, ,
\label{eq:chi}
\end{equation}
where $N_{x}$ represents the number of events evaluated by integrating Eq.~(\ref{eq:dNdTA}) over the
nuclear recoil energy.
Here, $N_{\text{SM}}^i$ refers to the number of events expected  in the $i$th bin according to the SM, while
$N_{\text{new}}^i$ includes an extra contribution associated to the relevant new physics of interest. The statistical uncertainty is defined as $\sigma^i_{\text{stat}}=\sqrt{N^i_{\text{SM}} + N^i_{\text{bg}}}$, while $\mathtt{a}$ denotes a total normalization factor handled as a nuisance parameter accounting for the systematic uncertainty, for which we employ two benchmark values:
$\sigma_{\text{sys}}=$ 2\% and 5\%.

For the case of the CENNS-610 detector, the background events are
  due to the beam related neutrons (BRN), whose energy dependence is
  provided by the CENNS-10 data release~\footnote{The data imply
      that $\sim 94$\% ($\sim 6$\%) correspond to prompt (delayed) BRN
      events. The same proportion will be used for the CENNS-610
      detector.}~\cite{Akimov:2020czh}. GEANT4 simulations performed
  by the COHERENT Collaboration in Ref.~\cite{Akimov:2019xdj}
  suggested that an improved background rate of 0.53 per kg.yr prompt
  BRN events is achievable at CENNS-610 with the aid of a neutron
  moderator. For 1 year of exposure this translates
  into 323 (21) prompt (delayed) BRN events per
  year and implies that the total BRN background accounts to about
  10\% of the \cevns signal. We finally note that given the absence
  of relevant data for ESS and CCM, we will adopt a single-bin
  $\chi^2$ treatment and we will furthermore assume that
  $N_\text{bkg}=10\% N_\text{SM}$.

\begin{figure}[!h]
\includegraphics[height=7.5cm,width= \textwidth]{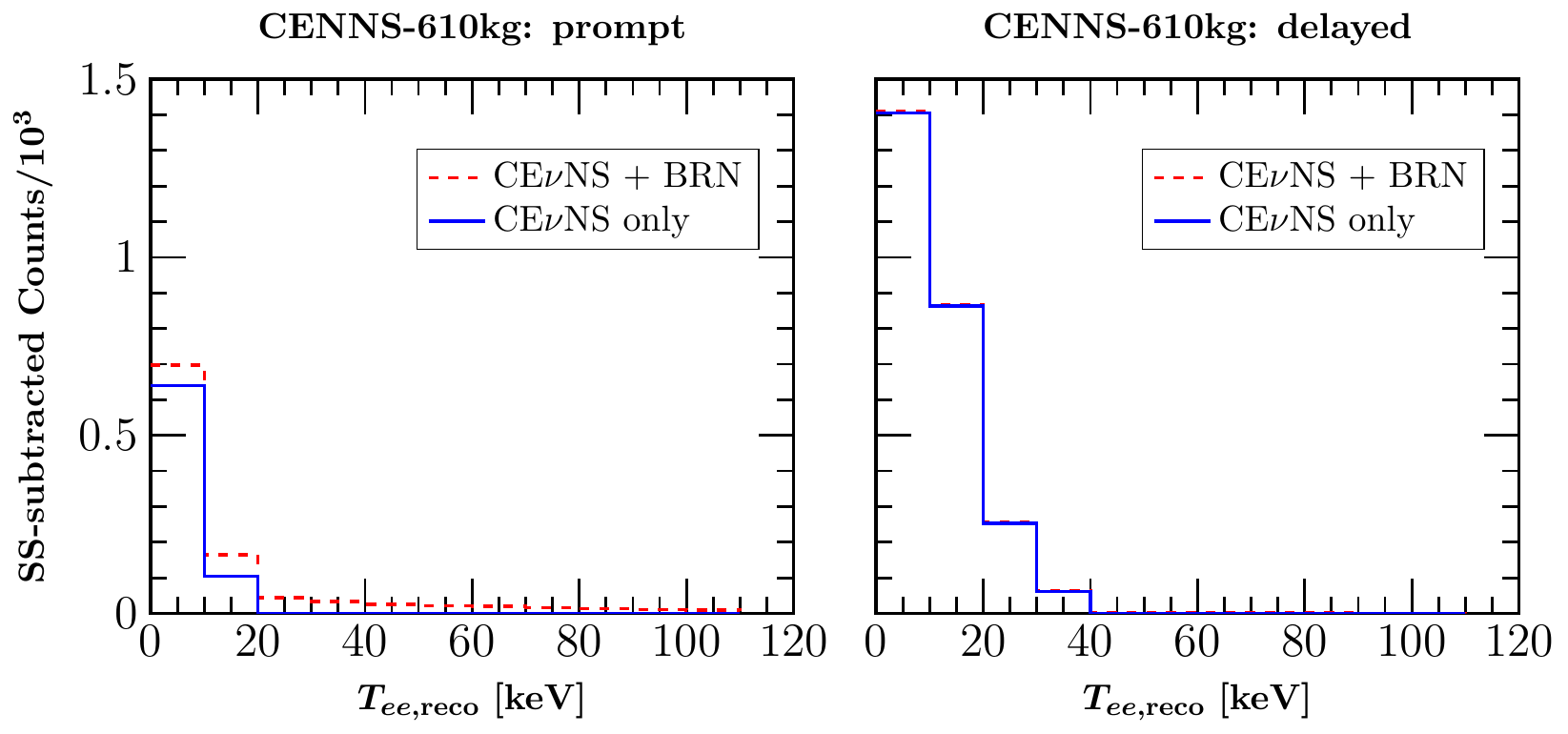}
\caption{Energy dependence of the expected CE$\nu$NS signal and BRN background assuming 1 year of data-taking  at the CENNS-610 experiment.}
\label{fig:signal}
\end{figure}

For the case of the CENNS-610 experiment, we calculate the number of events following the prescription described in the data release of the CENNS-10 experiment~\cite{Akimov:2020czh}. Thus, by using the quenching factor and the energy resolution,  we converted from nuclear recoil space $T_A$ $\mathrm{[keV_{nr}]}$ to the reconstructed electron equivalent energy space $T_{ee,\text{reco}}$ $\mathrm{[keV_{ee}]}$. We have furthermore verified the total of 128 CE$\nu$NS events predicted for the analysis A in Ref.~\cite{Akimov:2020pdx}, as well as confirmed that the energy dependence of our calculated CE$\nu$NS signal matches with Ref.~\cite{Akimov:2020pdx}. Then, we have appropriately scaled up our calculated number of events to account for the CENNS-610 specifications. In what follows, we will further assume that the steady-state (SS) background is well understood from beam-OFF measurements, and hence our calculations will involve the SS-subtracted signal. The spectral dependence of the expected signal after 1 year of data-taking at the CENNS-610 experiment is shown in Fig.~\ref{fig:signal} and compared with the BRN background for the prompt and delayed timing windows. 


\section{heavy singlet neutrinos and nonunitarity}
\label{sec:heavy-singl-neutr}

  Here we assume that, in addition to the three standard light neutrinos, one has extra singlet neutral heavy leptons that mediate light-neutrino mass generation.
It is well known that such heavy leptons will couple subdominantly in the weak charged current, via mixing with the SM isodoublet neutrinos~\cite{Schechter:1980gr}.
In the most general case, their presence and mixing with the active neutrinos respects the chiral SM structure.
Alternatively, we also consider the possibility of light sterile neutrinos taking part in oscillations.
  Both lead to new features beyond the minimal three-neutrino oscillation paradigm.
Here we note that constraining nonunitarity effects at short baselines plays a crucial role in mitigating the ambiguities present in testing for leptonic $CP$ violation in long-baseline neutrino oscillation experiments~\cite{Miranda:2016wdr}. We propose to do this through the neutral-current.

In this section, we consider \cevns experiments in the presence of unitarity violation effects. 
To set up notation, we write the relevant generalized charged-current weak interaction mixing matrix as
\begin{equation}
N= N^{\mathrm{NP}} U^{3 \times 3} \, ,
\end{equation}
where $U^{3 \times 3}$ denotes the standard unitary lepton mixing matrix and $N^{\mathrm{NP}}$ represents the new physics (NP) matrix which accounts for unitarity violation~\cite{Escrihuela:2015wra}. The latter is parametrized as 
\begin{equation}
N^{\mathrm{NP}}=
\left(
\begin{array}{lcccl} 
\alpha_{11} &    0        & 0  \\
\alpha_{21} & \alpha_{22} & 0  \\
\alpha_{31} & \alpha_{32} & \alpha_{33}\\
\end{array}\right) \, ,
\end{equation}
with the diagonal (off-diagonal) components $\alpha_{ii}$ ($\alpha_{ij}$) being real (complex) numbers.
In this context, the oscillation probability for $\nu_\alpha \to \nu_\beta$ transitions reads
\begin{equation}
\begin{aligned}
P_{\alpha \beta} =  \sum^3_{i,j} N^*_{\alpha i}N_{\beta i}N_{\alpha j}N^*_{\beta j} -& 
        4 \sum^3_{j>i} Re\left[
        N^*_{\alpha j}N_{\beta j}N_{\alpha i}N^*_{\beta i}\right] 
        \sin^2\left(\frac{\Delta m^2_{ji}L}{4E_\nu}\right)  \\
+ & 
        2 \sum^3_{j>i} Im\left[
        N^*_{\alpha j}N_{\beta j}N_{\alpha i}N^*_{\beta i}\right] 
        \sin\left(\frac{\Delta m^2_{ji}L}{2E_\nu}\right)  \, .
\end{aligned}
\end{equation}
The survival probabilities $P_{ee}$ and $P_{\mu \mu}$~\footnote{Here we neglect cubic products of small
  parameters $\alpha_{21}$, $\sin\theta_{13}$, and $\Delta m^2_{21}/\Delta m^2_{31}$.} and the transition probability
$P_{\mu e}$ simplify to~\cite{Escrihuela:2015wra}
\begin{equation}
\begin{aligned}
P_{ee} =& \alpha^4_{11} P_{ee}^{3\times3}\, , \\         
P_{\mu\mu} =& \alpha_{22}^4 P^{3\times3}_{\mu\mu} 
          + \alpha_{22}^3|\alpha_{21}|   P^{I_1}_{\mu\mu} 
          + 2|\alpha_{21}|^2\alpha_{22}^2 P^{I_2}_{\mu\mu}\, , \\
P_{\mu e} = & 
 (\alpha_{11}\alpha_{22})^2 P^{3\times3}_{\mu e}
+  \alpha_{11}^2 \alpha_{22}|\alpha_{21}|  P^{I}_{\mu e} 
+ \alpha_{11}^2|\alpha_{21} |^2 \, .
\end{aligned}           
\label{eq:prob}
\end{equation}
Here, $P^{3\times3}_{ee}$, $P^{3\times3}_{\mu\mu}$ and $P^{3\times3}_{\mu e}$ denote the standard oscillation probabilities, while the extra terms $P^{I_1}_{\mu\mu}$ and
$P^{I_2}_{\mu\mu}$ are defined in Ref.~\cite{Escrihuela:2015wra}.
Notice that $P^{I_1}_{\mu\mu}$ depends on a new $CP$ violation phase, $I_{NP}$, while $P^{I_2}_{\mu\mu}$  is phase independent. \\[-.2cm]

For the short-baseline \cevns experiments we are interested in here, there is no time for oscillations among active neutrinos to develop.
Hence, the baseline dependence in Eq. (\ref{eq:prob}) is not relevant~\footnote{For the $\pi$-DAR  \cevns
    experiments, $L=20-40$~m and $E_\nu \sim$~a few MeV; thus $\Delta m^2_{i1} L/E_\nu \ll 1$ for $i=2,3$.}.
Therefore, the effect of the heavy neutrino states at \cevns experiments will be mainly due to the zero-distance effect,
i.e. $P_{\alpha \beta}(L=0)$. The zero-distance probabilities are given as
\begin{eqnarray}
P_{ee} &=& \alpha_{11}^4 , \nonumber \\
P_{\mu\mu} &=&  (|\alpha_{21}|^2+\alpha_{22}^2)^2, \\
P_{\mu e} &=&  \alpha_{11}^2|\alpha_{21}|^2, \nonumber \\
P_{e \tau} &=& \alpha_{11}^2 |\alpha_{31}|^2 , \nonumber \\
P_{\mu  \tau} & \simeq & \alpha_{22}^2|\alpha_{32}|^2,  \nonumber
\label{eq:zero-dist-prob}
\end{eqnarray}
while the following ``triangle inequalities'' among the elements of the $N^\mathrm{NP}$ matrix hold~\cite{Fernandez-Martinez:2016lgt,Escrihuela:2016ube}
\begin{equation}
\begin{aligned}
|\alpha_{21}| \leq & \sqrt{(1-\alpha_{11}^2) (1-\alpha_{22}^2)} \, , \\
|\alpha_{31}| \leq & \sqrt{(1-\alpha_{11}^2) (1-\alpha_{33}^2)} \, , \\
|\alpha_{32}| \leq & \sqrt{(1-\alpha_{22}^2) (1-\alpha_{33}^2)} \, .
\end{aligned}
\label{eq:tringular_ineq}
\end{equation}

Within this context, due to the zero-distance effect,  neutrino fluxes at a spallation source are modified as follows:
\begin{equation}
\begin{aligned}
\frac{\mathrm{d}\phi_{e}^{\mathrm{NU}}}{\mathrm{d}E_\nu } = \frac{\mathrm{d}\phi_{\nu_e}^{\mathrm{NU}}}{\mathrm{d}E_\nu } + \frac{\mathrm{d}\phi_{\overline{\nu}_e}^{\mathrm{NU}}}{\mathrm{d}E_\nu }
 =& P_{ee} \, \frac{\mathrm{d}\phi_{\nu_e}^0}{\mathrm{d}E_\nu } + P_{\mu e} \left( \frac{\mathrm{d}\phi_{\nu_{\mu }}^0 }{\mathrm{d}E_\nu } +  \frac{\mathrm{d}\phi_{\overline{\nu} _{\mu }}^0 }{\mathrm{d}E_\nu } \right)  \, ,\\
  \frac{\mathrm{d}\phi_{\mu}^{\mathrm{NU}}}{\mathrm{d}E_\nu } =
\frac{\mathrm{d}\phi_{{\nu}_{\mu}}^{\mathrm{NU}}}{\mathrm{d}E_\nu } +
\frac{\mathrm{d}\phi_{\overline{\nu} _{\mu}}^{\mathrm{NU}} }{\mathrm{d}E_\nu } =& P_{e \mu} \,  \frac{\mathrm{d}\phi_{\nu_e}^0}{\mathrm{d}E_\nu } + P_{\mu \mu } \left( \frac{\mathrm{d}\phi_{\nu_{\mu }}^0 }{\mathrm{d}E_\nu } +  \frac{\mathrm{d}\phi_{\overline{\nu}_{\mu }}^0 }{\mathrm{d}E_\nu } \right)\, , \\
\frac{\mathrm{d}\phi_{\tau}^{\mathrm{NU}}}{\mathrm{d}E_\nu }  = 
\frac{\mathrm{d}\phi_{\nu_\tau}^{\mathrm{NU}}}{\mathrm{d}E_\nu } +
\frac{\mathrm{d}\phi_{\overline{\nu}_\tau}^{\mathrm{NU}}}{\mathrm{d}E_\nu } =& P_{e \tau} \, \frac{\mathrm{d}\phi_{\nu_e}^0}{\mathrm{d}E_\nu } + P_{\mu \tau} \, \left( \frac{\mathrm{d}\phi_{\nu_{\mu }}^0 }{\mathrm{d}E_\nu } +  \frac{\mathrm{d}\phi_{\overline{\nu} _{\mu }}^0 }{\mathrm{d}E_\nu } \right)   \, ,
\end{aligned}
\label{eq:NU-flux}
\end{equation}
where $\left(\frac{\mathrm{d}\phi_{\nu_e}^0}{\mathrm{d}E_\nu },
\frac{\mathrm{d}\phi_{\nu_{\mu }}^0 }{\mathrm{d}E_\nu },
\frac{\mathrm{d}\phi_{\overline{\nu}_{\mu }}^0 }{\mathrm{d}E_\nu
}\right)$ denote the unoscillated neutrino energy fluxes given in Eqs.(\ref{eq:PromptFlux})--(\ref{eq:DelFluxEl}).\\[-.2cm]

Most generally, the above expression can be written compactly as
\begin{equation}
  \label{eq:gen}
\begin{pmatrix}
\frac{\mathrm{d}\phi_{e}^{\mathrm{NU}}}{\mathrm{d}E_\nu }  \\
\frac{\mathrm{d}\phi_{\mu}^{\mathrm{NU}}}{\mathrm{d}E_\nu } \\
\frac{\mathrm{d}\phi_{\tau}^{\mathrm{NU}}}{\mathrm{d}E_\nu } 
\end{pmatrix} = 
\begin{pmatrix}
P_{ee} & P_{\mu e} & P_{\tau e} \\
P_{e \mu } & P_{\mu \mu} & P_{\tau \mu} \\
P_{e \tau } & P_{\mu \tau} & P_{\tau \tau} \\
\end{pmatrix}
\begin{pmatrix}
\frac{\mathrm{d}\phi_{\nu_e}^{0}}{\mathrm{d}E_\nu }  \\
\frac{\mathrm{d}\phi_{\nu_\mu}^{0}}{\mathrm{d}E_\nu } \\
\frac{\mathrm{d}\phi_{\nu_\tau}^{0}}{\mathrm{d}E_\nu } 
\end{pmatrix}
+
\begin{pmatrix}
\overline{P_{ee}} & \overline{P_{\mu e}} & \overline{P_{\tau e}} \\
\overline{P_{e \mu }} & \overline{P_{\mu \mu}} & \overline{P_{\tau \mu}} \\
\overline{P_{e \tau }} & \overline{P_{\mu \tau}} & \overline{P_{\tau \tau}} \\
\end{pmatrix}
\begin{pmatrix}
\frac{\mathrm{d}\phi_{\bar{\nu}_e}^{0}}{\mathrm{d}E_\nu }  \\
\frac{\mathrm{d}\phi_{\bar{\nu}_\mu}^{0}}{\mathrm{d}E_\nu } \\
\frac{\mathrm{d}\phi_{\bar{\nu}_\tau}^{0}}{\mathrm{d}E_\nu } 
\end{pmatrix} \, .
\end{equation}
with $\overline{P_{\alpha\beta}}=P(\overline{\nu}_\alpha \to \overline{\nu}_\beta)$.
Given appropriate choices for its entries, Eq.~(\ref{eq:gen}) holds for any neutrino experiment with an arbitrary type of neutrino source and no charge identification.
For the specific spallation case, only the initial $\nu_e$, $\nu_\mu$ and  $\overline{\nu}_\mu$ are nonvanishing, so we obtain the expressions in Eq.~(\ref{eq:NU-flux}).

Note that, since the experiments under study cannot distinguish neutrinos from antineutrinos, we combine both contributions in a flavor-dependent signal, as indicated in Eq.~(\ref{eq:NU-flux}). 
There, we have also assumed that neutrino and antineutrino oscillation probabilities are equivalent,  $P(\nu_\alpha \to \nu_\beta) = P(\overline{\nu}_\alpha \to \overline{\nu}_\beta) = P_{\alpha\beta}$ and also that $P_{\alpha \beta} = P_{\beta \alpha}$.
As seen from Eq.~(\ref{eq:NU-flux}), an additional monochromatic $\nu_e$ beam is generated  due to $\nu_\mu \to \nu_e$ transition, as well as a continuous $\bar{\nu}_e$ spectrum due to $\bar{\nu}_\mu \to \bar{\nu}_e$ conversion. Similarly, a new tau-neutrino flux is also expected due to $\nu_e \to \nu_\tau$, $\nu_\mu \to \nu_\tau$ and $\bar{\nu}_\mu \to \bar{\nu}_\tau$ oscillations.
However, one finds that these fluxes are largely suppressed due to the smallness of the appearance probabilities $P_{e\tau}$ and $P_{\mu\tau}$, well constrained by the existing limits on the
   nonunitarity (NU) parameters $\alpha_{ij}$.
\begin{figure}[t]
\includegraphics[width= 0.5\textwidth]{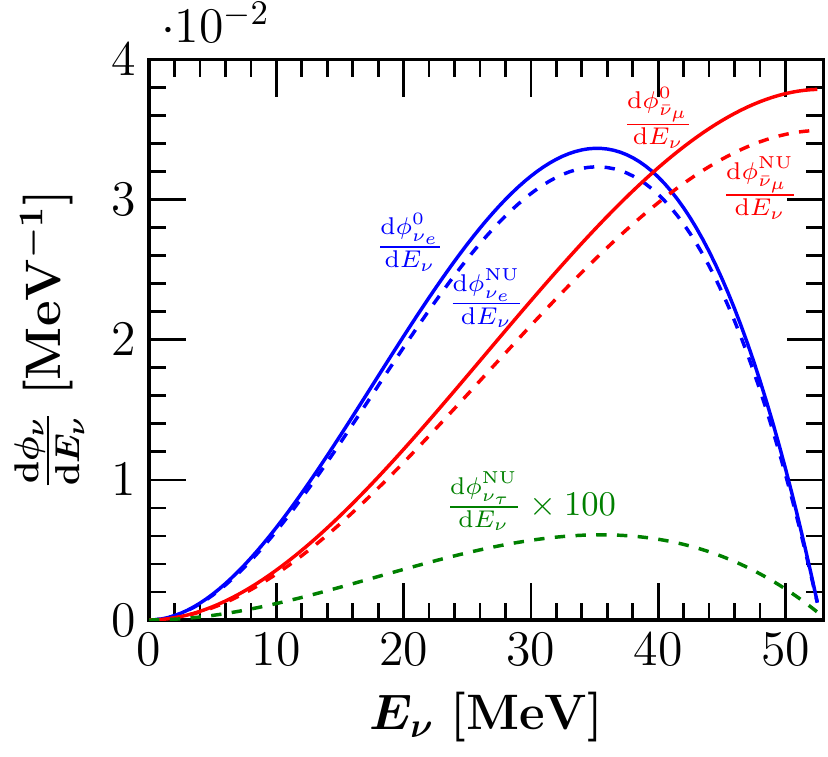}
\caption{ Flavor composition of the continuous $\pi$-DAR neutrino spectra in the SM (solid lines) and with
    nonunitarity effects (dashed lines), taking for these the maximal deviation parameters $\alpha_{ij}$ allowed at 90\% C.L.~\cite{Escrihuela:2016ube}.}
\label{fig:flux-NU}
\end{figure} 
 The flavor components of the corresponding continuous fluxes are displayed in Fig.~\ref{fig:flux-NU}.
In this figure, we show the modification of the initial neutrino flux due to the zero-distance nonunitarity effect.
The modified spectra have been evaluated using the 90\% C.L. limits on the $\alpha_{ij}$ parameters reported in Ref.~\cite{Escrihuela:2016ube}. 

  In what follows, we give a first estimate on the prospects for probing the unitarity-violating parameters at future liquid argon detectors.
  In order to determine the sensitivity limits on unitarity violation, we proceed as explained in Ref.~\cite{Escrihuela:2015wra}.
  For definiteness, we will focus on the detection of electron and muon neutrinos, reducing the number of relevant NU parameters to three: $\alpha_{11}$, $\alpha_{22}$ and
  $|\alpha_{21}|$~\footnote{Note that the nondiagonal parameter $\alpha_{21}$ is complex, but the zero-distance effects analyzed here are insensitive to the associated phase.}. \\[-.3cm]

\begin{figure}[t]
\includegraphics[width=  \textwidth]{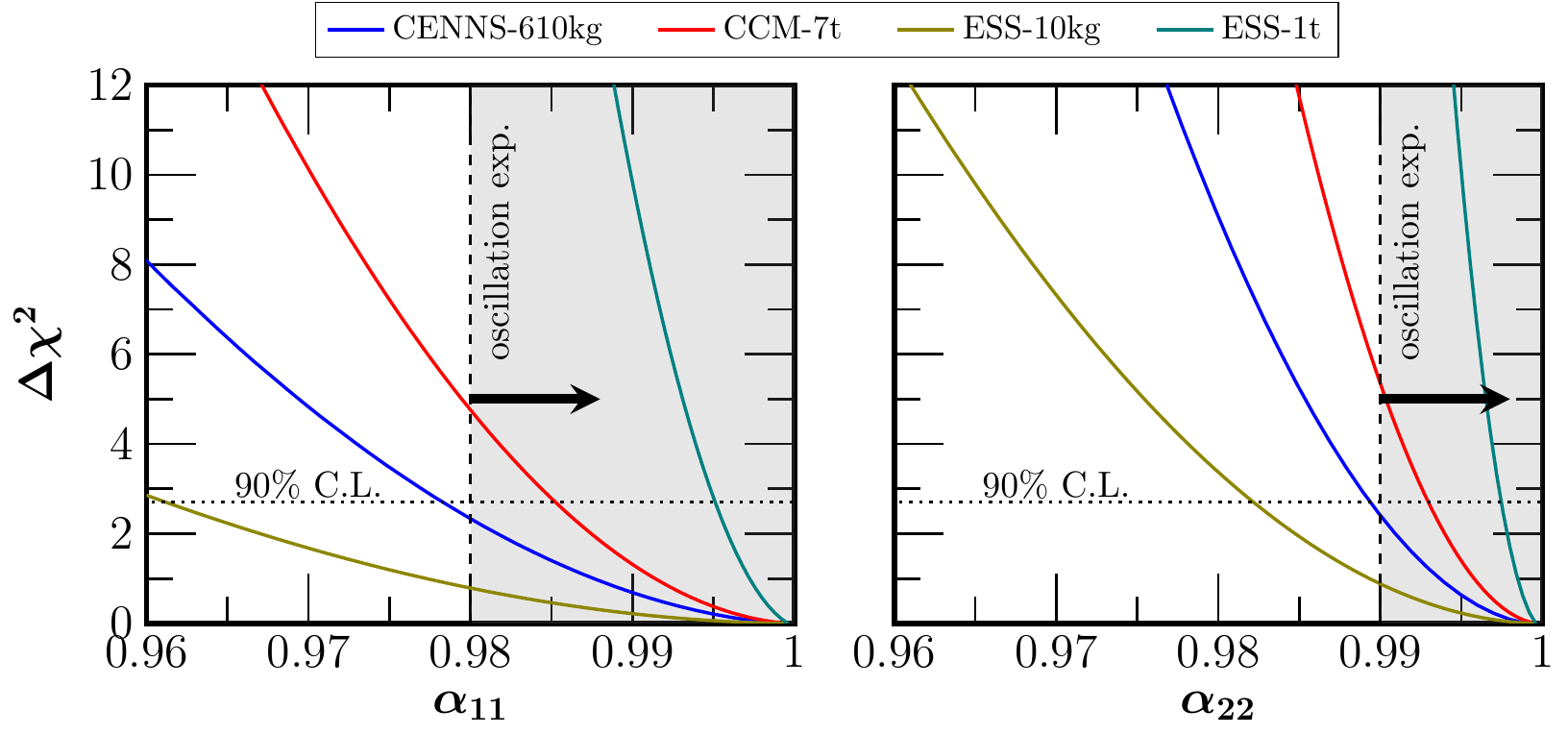}
\caption{Sensitivity on the diagonal parameters $\alpha_{11}$ (left) and $\alpha_{22}$ (right), marginalized over the undisplayed parameters for different experimental configurations.
  For comparison we also give the sensitivity obtained from global oscillation data analysis~\cite{Escrihuela:2016ube}.}
\label{fig:chi-diagonal}
\end{figure}
\begin{figure}[ht]
\begin{center}
\includegraphics[width= 0.49 \textwidth]{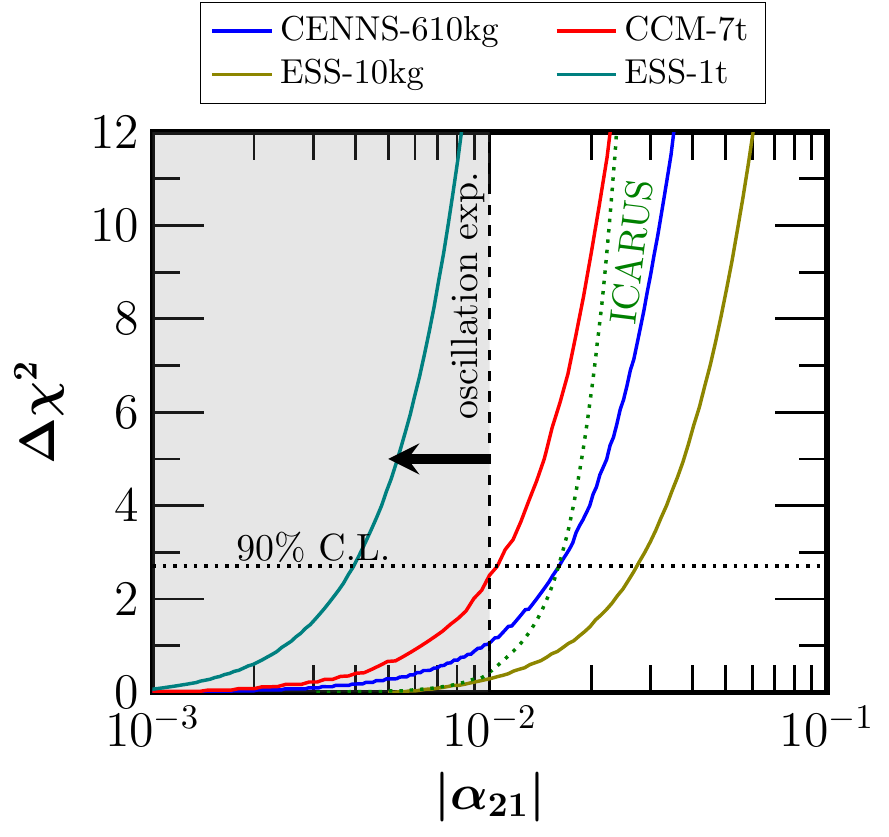}
\includegraphics[width= 0.45 \textwidth]{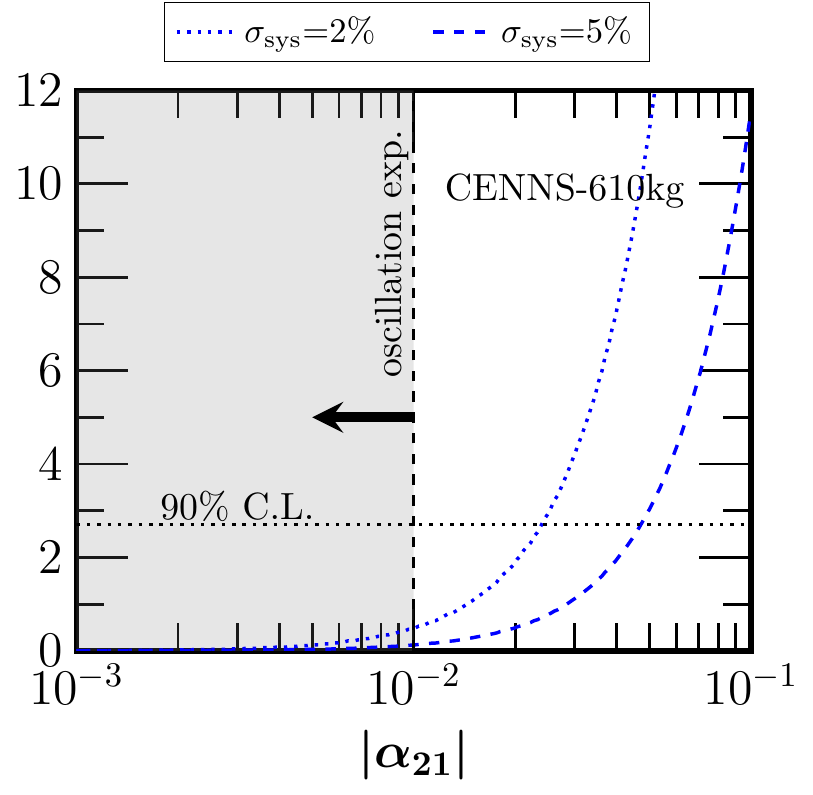}
\end{center}
\caption{Sensitivity on the nondiagonal parameter $|\alpha_{21}|$  marginalized over the undisplayed parameters for different experimental configurations. 
A comparison with the analysis of ICARUS data~\cite{Miranda:2018yym} as well as global oscillation data~\cite{Escrihuela:2016ube} is also given. The left panel shows the results neglecting systematic uncertainties, while the right panel takes into account systematic uncertainties for the case of CENNS-610. }
\label{fig:chi2-nondiagonal}
\end{figure}

\begin{table}[t]
\begin{tabular}{|c|l|c|c|c|}
\hline
\multicolumn{2}{|c|}{Experiment}                                                                                           & \multicolumn{1}{c|}{$\alpha_{11}$}                                             & \multicolumn{1}{c|}{$\alpha_{22}$}                                             & \multicolumn{1}{c|}{$|\alpha_{21}| \, (\times 10^{-2})$}                        \\ \hline
\multicolumn{2}{|c|}{Oscillations~\cite{Escrihuela:2016ube}}                                                                                 & > 0.98                                                                         & > 0.99                                                                         & < 1.0                                                                  \\ \hline
 $\sigma_\text{sys} = 0$                                           & \begin{tabular}[c]{@{}l@{}}CENNS-610kg\\ CCM-7t\\ ESS-10kg\\ ESS-1t \end{tabular} & \begin{tabular}[c]{@{}l@{}}> 0.978\\ > 0.985\\ > 0.961 \\ > 0.995\end{tabular} & \begin{tabular}[c]{@{}l@{}}> 0.989\\ > 0.993\\ > 0.982 \\ > 0.997\end{tabular} & \begin{tabular}[c]{@{}l@{}}< 1.6\\ < 1.1\\ < 2.7 \\ < 0.4\end{tabular} \\ \hline
 $\sigma_\text{sys} = 2\%$                                        & \begin{tabular}[c]{@{}l@{}}CENNS-610kg\\ CCM-7t\\ ESS-10kg\\ ESS-1t \end{tabular} & \begin{tabular}[c]{@{}l@{}}> 0.967\\ > 0.971\\ > 0.954 \\ > 0.976\end{tabular} & \begin{tabular}[c]{@{}l@{}}> 0.984\\ > 0.986\\ > 0.979 \\ > 0.988\end{tabular} & \begin{tabular}[c]{@{}l@{}}< 2.4\\ < 2.1\\ < 3.3 \\ < 1.7\end{tabular} \\ \hline
 $\sigma_\text{sys} = 5\%$                                       & \begin{tabular}[c]{@{}l@{}}CENNS-610kg\\ CCM-7t\\ ESS-10kg\\ ESS-1t \end{tabular} & \begin{tabular}[c]{@{}l@{}}> 0.933\\ > 0.934\\ > 0.924\\ > 0.937\end{tabular}  & \begin{tabular}[c]{@{}l@{}}> 0.969\\ > 0.970\\ > 0.966 \\ > 0.971\end{tabular} & \begin{tabular}[c]{@{}l@{}}< 4.7\\ < 4.6\\ < 5.3 \\ < 4.5 \end{tabular} \\ \hline
\end{tabular}
\caption{90\% C.L. sensitivities on unitarity deviations from our present analysis of liquid argon \cevns experiments.
    We also give a comparison with results from the global neutrino oscillation data analysis~\cite{Escrihuela:2016ube}.}
\label{tab:limits}
\end{table}%

\begin{figure}[h]
\includegraphics[width= \textwidth]{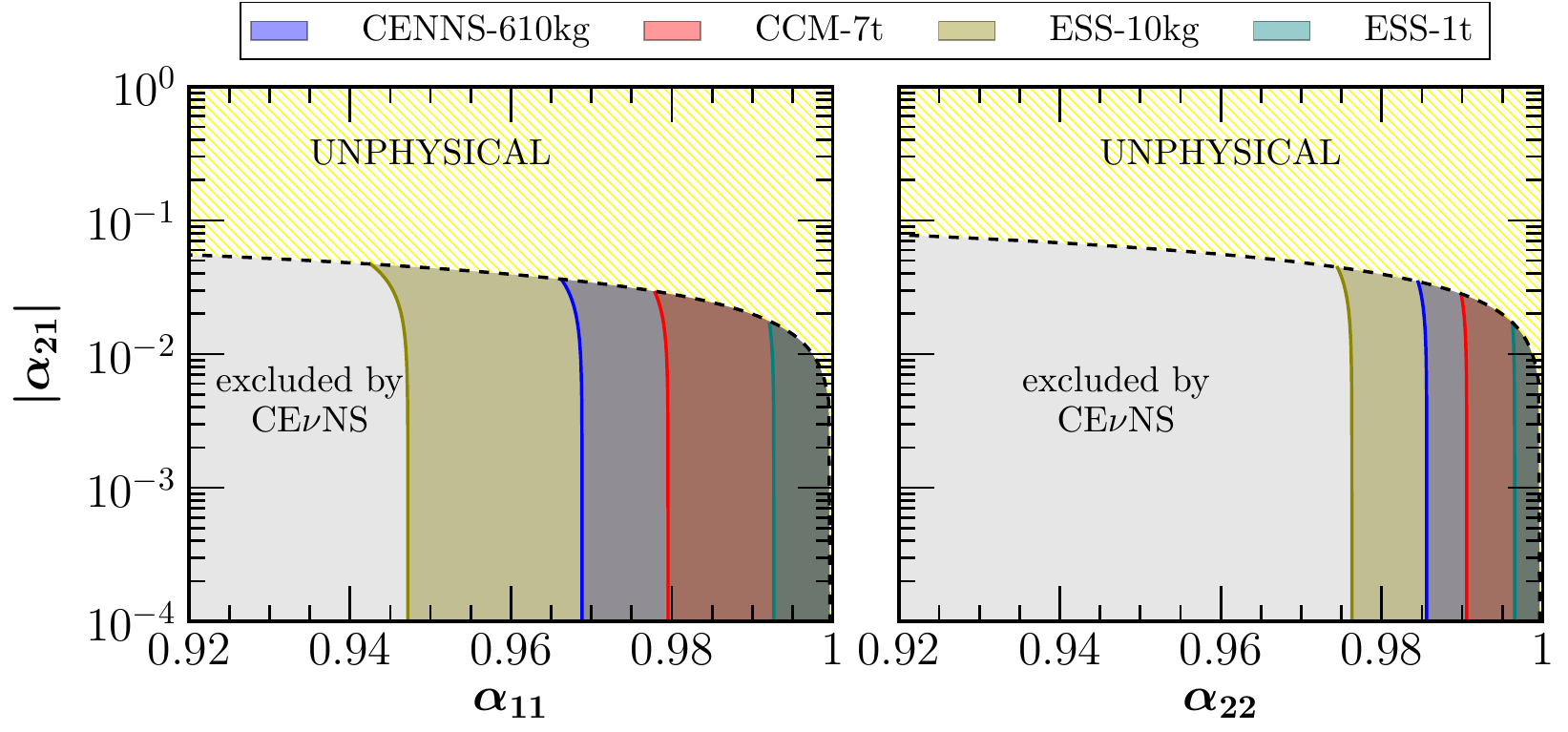}\\
\includegraphics[width= \textwidth]{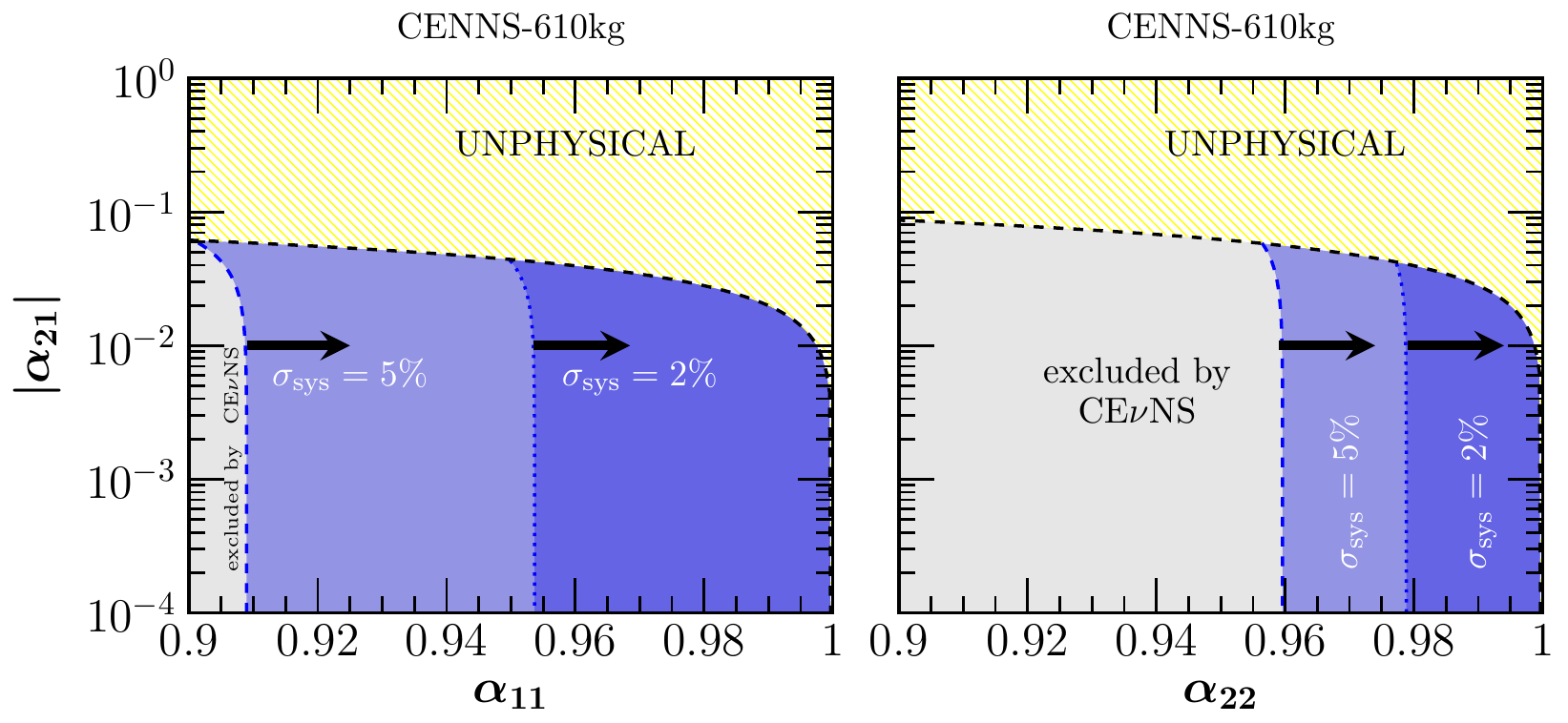}\\
\caption{Allowed regions at 90\% C.L. in the planes $\alpha_{11}$--$|\alpha_{21}|$ (left) and $\alpha_{22}$--$|\alpha_{21}|$ (right), marginalized over the undisplayed parameter. The gray shaded area below the black dashed curve denotes the bound given by Eq.~(\ref{eq:tringular_ineq}), while the yellow region above corresponds to the unphysical area.
Upper panel: the blue (red) [olive] \{teal\} shaded area corresponds to the analysis of the CENNS (CCM) [ESS-10kg] \{ESS-1t\} experiment, considering $\sigma_{sys} = 0$ in all cases.
Lower panel: CENNS-610 analysis  taking into account systematic uncertainties.}
  \label{fig:contour-NU}  
\end{figure}

Using the $\chi^2$ function defined in Eq.~(\ref{eq:chi}), we first perform our statistical analysis on the sensitivity to unitarity-violating parameters
$\alpha_{ij}$ by assuming $\sigma_\text{sys} = 0$. We varied only one parameter, marginalizing over the other two, and imposing the constraint coming from the triangular inequality of Eq. (\ref{eq:tringular_ineq}).
The ``one-at-a-time'' sensitivity profiles of future \cevns experiments for the diagonal parameters $\alpha_{11}$ and $\alpha_{22}$ are shown in Fig.~\ref{fig:chi-diagonal}. 
Comparing these sensitivities with those derived from global neutrino oscillation data~\cite{Escrihuela:2016ube},
one sees that the \cevns experiments might eventually become competitive with current oscillation searches.
Indeed, while the current configuration of ESS with 10~kg detector mass is not expected to be competitive, the next-generation of ESS will certainly have the
capability of improving current oscillation sensitivities, provided the systematic uncertainties are under control (see the discussion below).
In the left panel of Fig.~\ref{fig:chi2-nondiagonal} we illustrate the sensitivities on the modulus of the nondiagonal parameter $\alpha_{21}$, neglecting systematic uncertainties.
Our results are compared with upper limits obtained from global oscillation fits~\cite{Escrihuela:2016ube} and with the sensitivity of future ICARUS data, as estimated in Ref.~\cite{Miranda:2018yym}.
Concerning the prospects on the $|\alpha_{21}|$ sensitivity, one sees that most \cevns experiments cannot compete with current bounds.
However, the future ESS configuration may offer the chance of improving this situation drastically.
The right panel of Fig.~\ref{fig:chi2-nondiagonal} illustrates the projected sensitivities on $|\alpha_{21}|$ for the case of CENNS-610
taking into account systematic uncertainties of  2\%  and 5\%. This reduces the sensitivity by a factor $\sim2$  and $\sim4$, respectively, compared to the ideal case with $\sigma_\text{sys}=0$.
Similar conclusions hold for CCM and ESS.
Likewise, we have also checked that systematic uncertainties substantially diminish the corresponding sensitivities on $\alpha_{11}$ and $\alpha_{22}$.
A summary of the bounds we extract is given in Table~\ref{tab:limits}. As expected, one finds that a better control of the background events and systematic uncertainties will lead to improved sensitivities. For comparison, the current upper bounds derived from oscillation searches~\cite{Escrihuela:2016ube} are also given in Table~\ref{tab:limits}.

One can also perform a combined $\chi^2$ analysis through a simultaneous variation of two NU parameters, and marginalizing over the third one.
Our results for the CENNS, CCM and ESS experiments (current as well as next-generation setups) are presented in the upper panel of Fig.~\ref{fig:contour-NU}.
For each \cevns experiment, the dark-shaded areas in the $\alpha_{11}-|\alpha_{21}|$ and $\alpha_{22}-|\alpha_{21}|$ planes
located to the right of the lines are allowed at 90\% C.L. by the corresponding experiments. 
The region consistent with the triangle inequality of Eq.~(\ref{eq:tringular_ineq}) is the one below the dashed line in both panels. 
Therefore, the allowed values in the $\alpha_{ii}-|\alpha_{12}|$ plane are eventually determined by the intersection of the gray shaded area with the allowed region determined by each experiment's sensitivity.
  We find that CENNS and CCM have the potential to probe part of the currently allowed parameter space. 
As before, the most promising experimental setup is provided by the next phase of ESS with a ton-scale detector.
The lower panel of Fig.~\ref{fig:contour-NU} shows the allowed parameter space from the analysis of CENNS-610 when assuming systematic uncertainties $\sigma_\text{sys}=2\%$ and $\sigma_\text{sys}=5\%$.
The results imply that the determination of the diagonal NU parameters $\alpha_{ii}$ is more sensitive to the systematic uncertainties than that of the nondiagonal ones.
Improvements with respect to the present analysis could be obtained through a time-dependent study which allows reducing the background/signal ratio~\cite{Dutta:2019eml, Coloma:2019mbs}.

\section{light sterile neutrinos in (3+1) scheme}
\label{sec:light-ster-neutr} 

\begin{figure}[t]
\includegraphics[width= 0.4\textwidth]{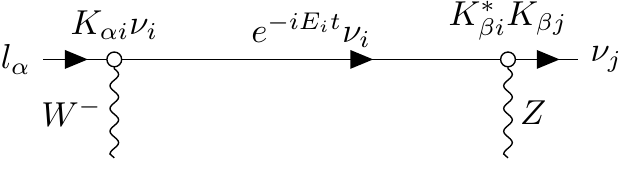}\\
\caption{Feynman diagram representing the charged-current production,
  followed by oscillation and neutral-current detection. There is a sum
  over the subindex $\beta$.}
\label{fig:NCosc}  
\end{figure}

Though the theoretical motivation is not especially strong, there could well be singlet neutrinos in nature, light enough to take part in oscillations, usually known as light sterile neutrinos.
 Although this situation differs from what we have considered above, it can be described within the same formalism developed in Ref.~\cite{Schechter:1980gr}. Here we present basically the same reasoning in a somewhat more modern form.
  There is a basic difference compared to most neutrino oscillation experiments, in which neutrinos are produced and detected through the charged-current (CC) weak interaction.
  Here neutrinos are produced conventionally, but detected through the neutral-current, as illustrated in Fig.~\ref{fig:NCosc}.
  The other important difference is that, since we cannot identify neutrino flavors, the process of interest is necessarily inclusive, with the observable being simply the recoil of the
  relevant nucleus.

For definiteness, we take the simplest (3+1) scheme with three active neutrinos $\nu_\alpha$ ($\alpha=e,\mu,\tau$) and one light sterile neutrino.
The overall quantum-mechanical amplitude for the process of interest is given as
\begin{equation}  
  \label{eq:amplitude}
{\cal{A}}_{\alpha j} = \sum^{4,3}_{i,\beta} K_{\alpha i}e^{-iE_i t}K^*_{\beta i}K_{\beta j} \, ,
  \end{equation}
  where the initial flavor index $\alpha$ is fixed, while $\beta$ is summed over the three flavors, and the roman (neutrino) mass index is summed from 1 to 4.
  One sees that, in the production CC vertex, one has the rectangular lepton mixing matrix $K$, then  the evolution factor,
\footnote{Due to the very short baseline, oscillations cannot develop and matter effects may be neglected.} and finally the NC detection vertex
  characterized by the projective matrix $P=K^\dagger K = P^2 = P$.
  Assuming the charged leptons are in their diagonal basis we can identify $K$ with the truncation of the $4\times 4$ unitary matrix $U$ diagonalizing the neutrinos,
  so the active flavors are expressed in terms of the four mass eigenstate neutrinos $\nu_i$ ($i= 1,2,3,4$) as $\nu_\alpha =\sum_i^4 U_{\alpha i} \nu_i$.
From this equation, we see that the survival probability to active neutrinos, $P_\alpha = \sum_\beta^3 P_{\alpha\beta}$, is given as 
\begin{eqnarray}
P_{\alpha} & = & \sum_{i,l,j,\beta,\beta'} K_{\alpha i}e^{-iE_i t}K^*_{\beta i}K_{\beta j}K^*_{\beta ' j}K_{\beta' l}e^{iE_l t}K^*_{\alpha l} \nonumber \\ 
&=& \sum_{i,l,\beta\beta'} K_{\alpha i}e^{-iE_i t}K^*_{\beta i}\delta_{\beta\beta'}
K_{\beta' l}e^{iE_l t}K^*_{\alpha l}  \\ \nonumber
&=& \sum_{i,l,\beta} K_{\alpha i}e^{-iE_i t}K^*_{\beta i} K_{\beta l}e^{iE_l t}K^*_{\alpha l}  \, ,
\end{eqnarray}
where greek indices run up to 3 and latin ones up to 4. This result corresponds to Eq.~(4.13) in Ref.~\cite{Schechter:1980gr}. Taking into account that the propagation factors are too small for the distances under consideration, except when the light sterile neutrino, corresponding to $i = 4$, is involved, we have 

\begin{eqnarray}
  \label{eq:prob4x4}
P_{\alpha} 
&\simeq & \sum_{l,\beta} K_{\alpha 4} K^*_{\beta 4} K_{\beta l} K^*_{\alpha l} e^{-i(E_4 -E_l) t}= \sum_{l,\beta} K_{\alpha 4} K^*_{\beta 4} K_{\beta l} K^*_{\alpha l} e^{-i\frac{\Delta m^2_{4l} L}{2E}}\, .
\end{eqnarray}
Notice that, as explained above, the active survival probability  $P_\alpha$ includes all the weak neutrino flavor states, $\nu_\beta$. 
 Note also that we can neglect the ``appearance'' part of this probability (i.e. the sum over the final $\nu_\mu$ and $\nu_\tau$ states for the case of an initial $\nu_e$), in comparison with the ``survival'' $\nu_e$ contribution.
  Indeed, the appearance probabilities will involve products of the form $\sin^2\theta_{i4} \sin^2\theta_{j4}$, and will be more suppressed than the ``survival'' part, that goes as $\sin^2\theta_{14}$. 
Hence, the above expression will lead to the usual vacuum survival probability
 \begin{equation}
   P_{ee}(E_\nu)  \simeq 1 - \sin^2 2\theta_{1 4} \sin^2 \left( \frac{\Delta m^2_{41} L}{4E_\nu}\right)\, ,
   \label{eq:prob_nues}
 \end{equation}
and similarly for  muon neutrinos
 \begin{equation}
 P_{\mu \mu}(E_\nu)  \simeq  1 - \sin^2 2\theta_{2 4} \sin^2 \left( \frac{\Delta m^2_{42} L}{4E_\nu}\right)\, , 
 \label{eq:prob_nums}
 \end{equation}
  with $\theta_{14},~\theta_{24}$ being the mixing angles and $\Delta m^2_{41} \approx \Delta m^2_{42}$ the mass splittings.
  The presence of the sterile neutrino is taken into account in the \cevns process through the substitution $\mathcal{Q}_W \to
   \mathcal{Q}_W P_{\alpha \alpha}(E_\nu) $ in the SM weak charge of Eq.~(\ref{eq:Qw}). 

We will now estimate the sensitivity of future \cevns experiments to the light sterile neutrino scenario.  To do this, we will use the
  formalism described in previous sections, but replace the neutrino oscillation probabilities in Eq.~(\ref{eq:NU-flux}) by the
  expressions in Eqs.~(\ref{eq:prob_nues}) and (\ref{eq:prob_nums}) above.
  Note that, unlike the case of nonunitarity, here oscillation probabilities depend on the neutrino energy.
  Our treatment of this scenario will also be slightly different, and we will consider independently oscillations in the channel $\nu_e \to \nu_s$ and $\nu_\mu \to \nu_s$.

\begin{figure}[t]
\includegraphics[width=0.48 \textwidth]{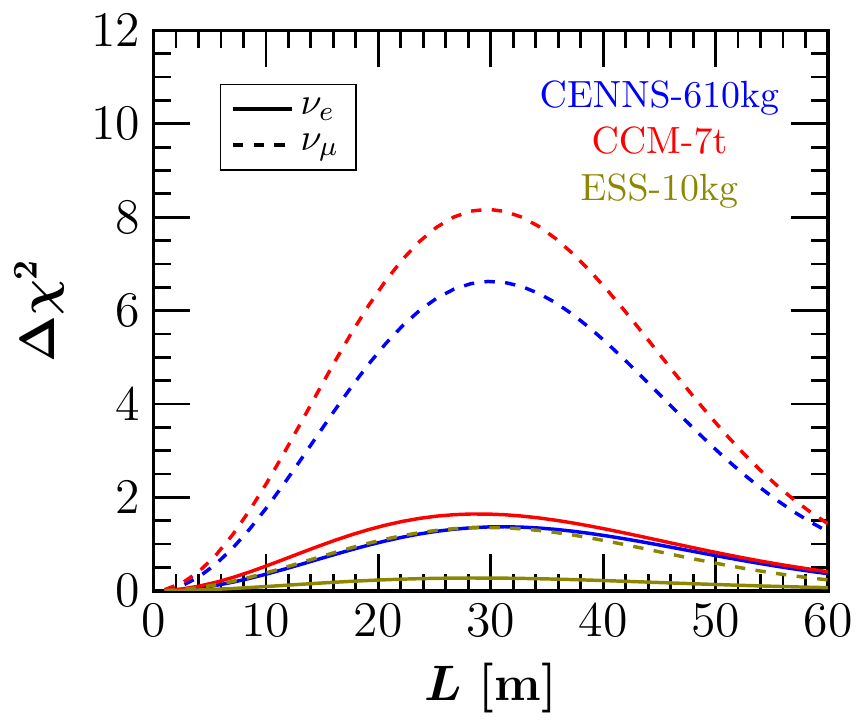}
\includegraphics[width=0.48 \textwidth]{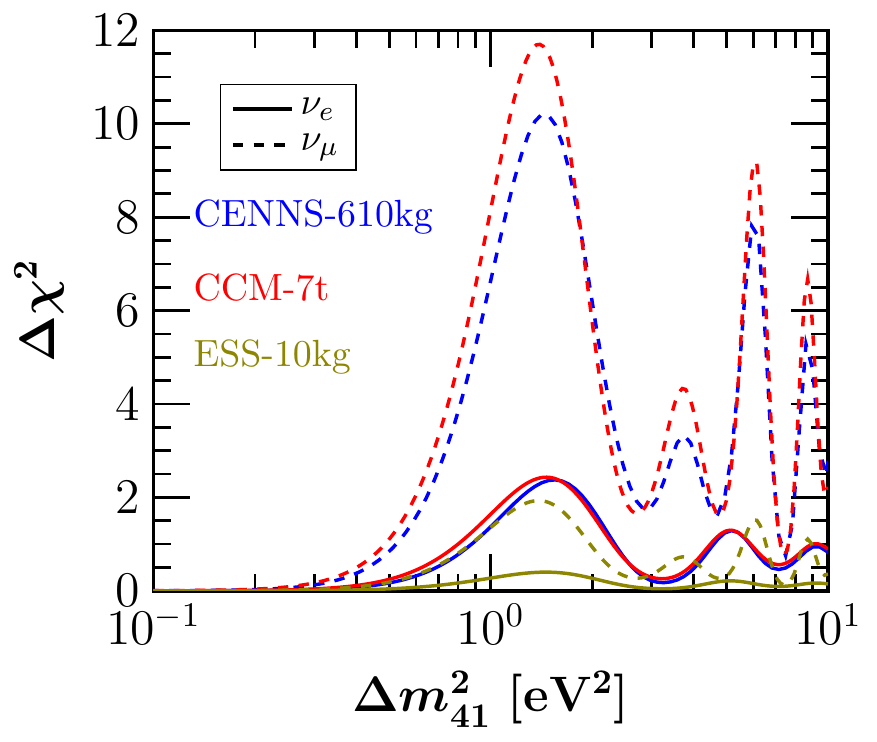}
\caption{Left panel: sensitivity profile with respect to the baseline $L$ for a fixed mass splitting $\Delta m_{41}^2=1~\mathrm{eV^2}$ for $\sin^2 2 \theta_{14}=0.1$ or $\sin^2 2
  \theta_{24}=0.1$ when dealing with $\nu_e$ (solid lines) or $\nu_\mu$ (dashed lines), respectively.  Blue (red) [olive] curves correspond to the CENNS (CCM) [ESS-10kg] experiment.
  Right-panel: sensitivity profile for the mass splitting $\Delta m_{41}^2$ for a fixed baseline $L=30$~m and fixed $\sin^2 2 \theta_{14}$ and $\sin^2 2\theta_{24}$ as in the left panel.}
\label{fig:baseline-mass}
\end{figure}

As a  first step, we explore the optimal baseline for light sterile neutrino searches with \cevns detectors.
For this purpose, we fix the sterile neutrino mixing parameters to the benchmark values $\Delta m^2_{41}=1 \mathrm{eV^2}$,
$\sin^2 2\theta_{14} = 0.1$ and $\sin^2 2\theta_{24} = 0.1$, and we evaluate the $\chi^2$ as a function of the baseline $L$.
The results obtained from the analysis of the CENNS, CCM and ESS experiments are shown in the left panel of Fig.~\ref{fig:baseline-mass}. 
As discussed before, we estimate independently the sensitivity for the electronic and muonic channel.
In all cases, the maximum sensitivity is reached around $L=30$~m,  very close to the proposed baselines. 
One sees how the CENNS and CCM experiments have the best sensitivity.
One can also note the larger sensitivity to sterile searches in the $\nu_\mu \to \nu_s$ channel in comparison with $\nu_e \to \nu_s$.
  Note, however, that to distinguish between these two oscillations channels, timing information would be required~\cite{Blanco:2019vyp}. 

We also find it useful to examine the sensitivity of the \cevns experiments to the mass splittings $\Delta m^2_{41}$. In the right
panel of Fig.~\ref{fig:baseline-mass} we illustrate the corresponding $\chi^2$ profiles by fixing $\sin^2 2\theta_{14} = 0.1$ or
$\sin^2 2\theta_{24} = 0.1$ and the baseline to $L=30$~m. As previously, CENNS and CCM perform better, while significantly higher sensitivities are reached when muon neutrinos are involved.
This is due to the larger flux of muon-like events emitted at spallation sources.
One also sees that, for our chosen mixing angle benchmarks, the $\Delta m_{41}^2$ mass splitting values for which one has better sensitivity are $1.5$ and $6~\mathrm{eV^2}$.
\begin{figure}[h]
\includegraphics[width=0.48 \textwidth]{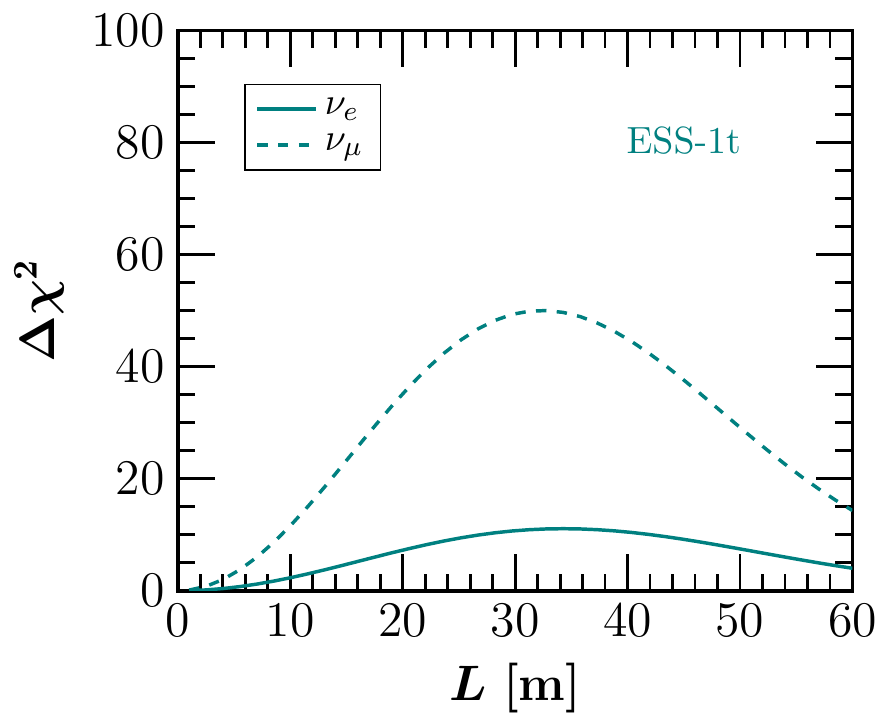}
\includegraphics[width=0.48 \textwidth]{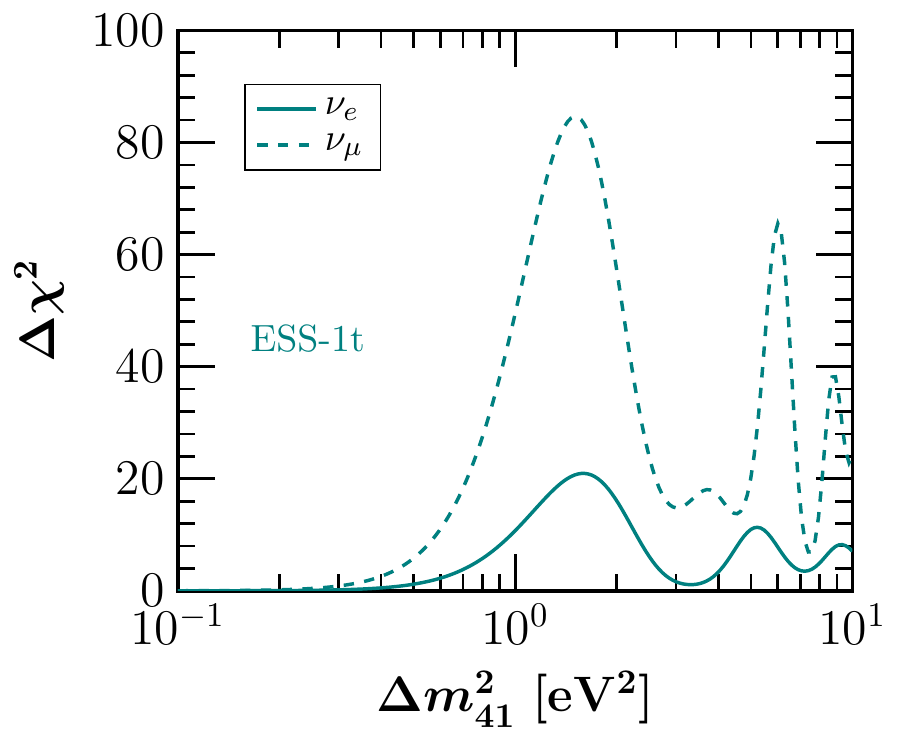}
\caption{Same as Fig.~\ref{fig:baseline-mass}, but for the future ESS with a 1 ton detector and 20 keV threshold.}
\label{fig:baseline-mass_future}
\end{figure}
\begin{figure}[t]
\includegraphics[width= \textwidth]{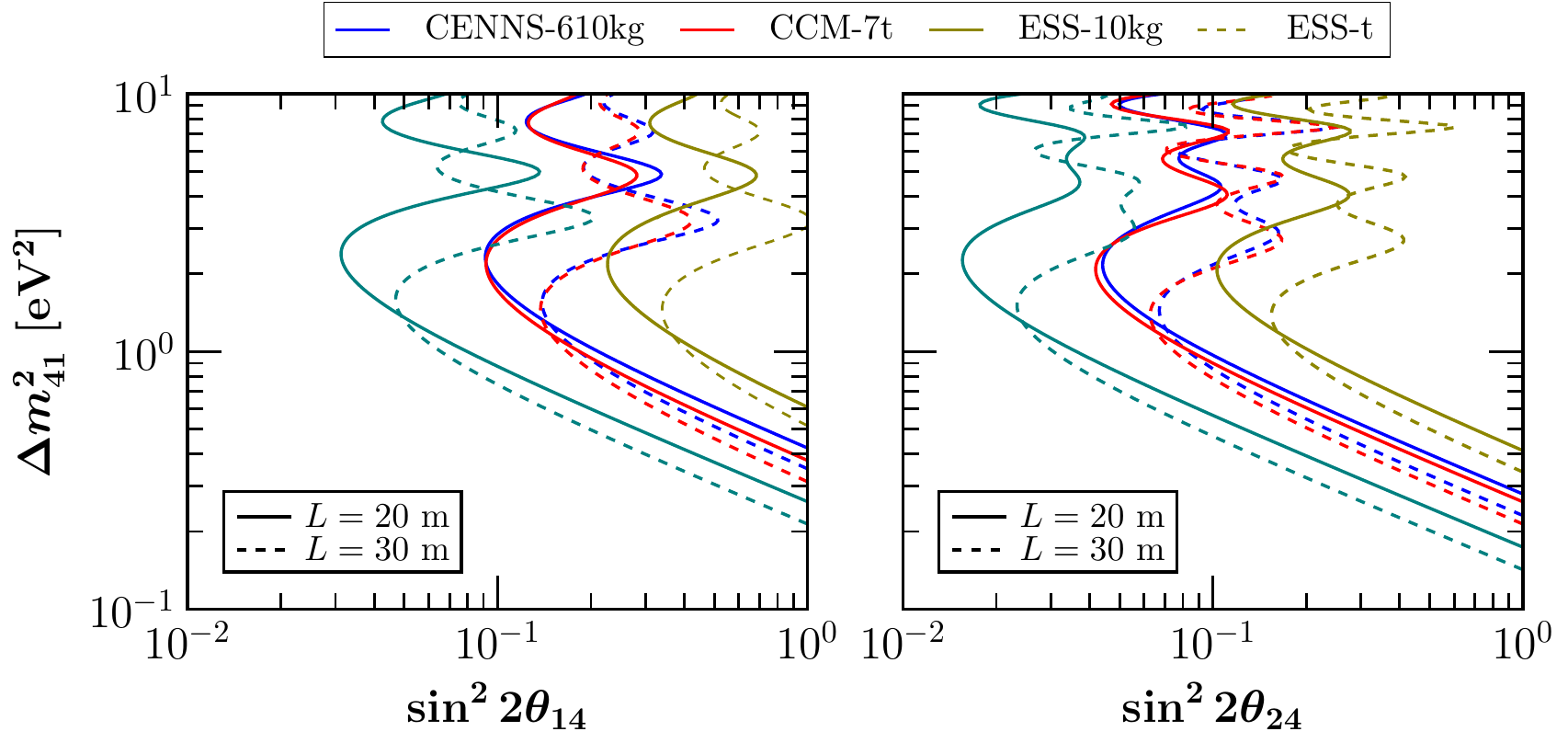}
\includegraphics[width= \textwidth]{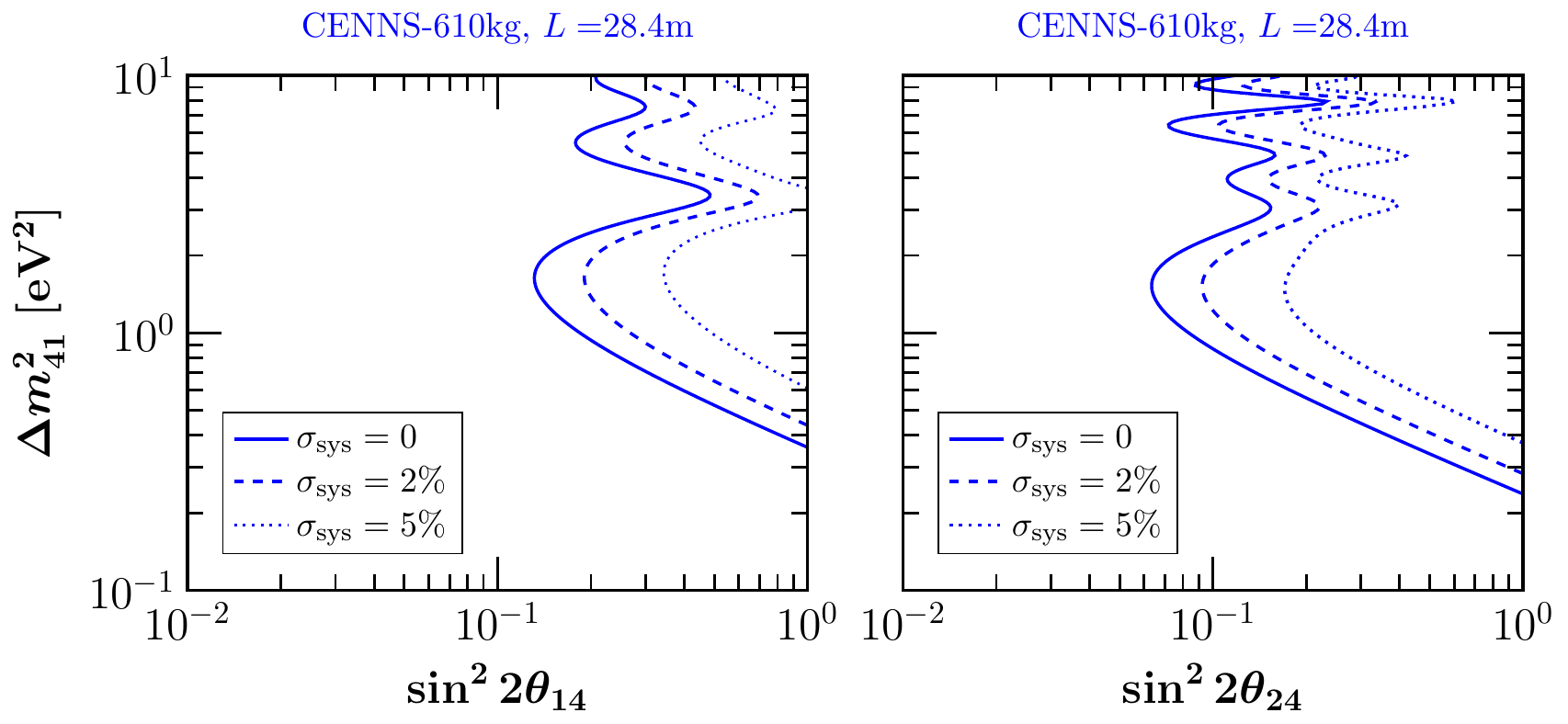}
\caption{Upper panel: 90\% C.L. sensitivity curves  in the $\sin^2 2 \theta_{i4}-\Delta m_{41}^2$ plane for different baselines and experiments with $\sigma_{sys} = 0$.
  Lower panel: CENNS-610 sensitivity to the sterile neutrino parameters for  $\sigma_{sys} =0, \,  2$ and  $5\%$ and a fixed baseline of $L=28.4$~m.}
\label{fig:contour}
\end{figure}

The attainable sensitivities of CENNS and CCM are very similar, despite the large difference with respect to their active detector masses.
Indeed, the highly intense neutrino flux available at the SNS can compensate the gain in exposure due to the large detector of CCM (see Table~\ref{tab:exps}).
On the other hand, the results obtained for the current configuration of ESS with a 10~kg detector mass are
promising, yet not competitive with the latter two since the detector size in this case is smaller by 2--3 orders of magnitude. However, ESS
offers the most intense neutrino beam, motivating us to perform an alternative analysis regarding its
future configuration with a 1 ton detector mass and a 20~keV threshold.
As illustrated in the left and right panels of Fig.~\ref{fig:baseline-mass_future}, ESS-based sterile neutrino searches are expected to be very promising in the long run.
Indeed, the highly intense neutrino beam available at the ESS can yield a very large number of events, making \cevns very relevant for short-baseline oscillation searches.

We now explore how the sterile neutrino parameter space can be probed via \cevns measurements at future large liquid argon detectors.
In our analysis, we vary simultaneously the mixing angle $\sin^2 2\theta_{i4}$ and the mass splitting $\Delta m^2_{41}$ = $\Delta m^2_{42}$, for different baselines.
The sensitivity curves at 90\% C.L. for the different experimental proposals considered in our study are presented in the left panel of Fig.~\ref{fig:contour}, for a vanishing systematic uncertainty. Focusing on CENNS-610, the right panel of Fig.~\ref{fig:contour} shows the modification of the allowed regions when the systematic uncertainties $\sigma_\text{sys}= 2\%$ and $\sigma_\text{sys}= 5\%$ are taken into account.
The results are rather promising, with the same general conclusions regarding the relative performance of the studied experiments.
We stress that the future configuration of the ESS experiment can become competitive with current precision oscillation studies.
  Indeed, our results illustrate the potential of neutral-current measurements in probing the parameter space constrained by global sterile-neutrino analyses;
  see e.g. Refs.~\cite{Gariazzo:2017fdh, Dentler:2018sju}.

\section{Summary and Outlook}
\label{sec:conclusions}

We have analyzed the potential of future \cevns experiments in probing new physics phenomena associated to the presence of heavy isosinglet neutrinos and light sterile neutrinos.
The purely neutral character of \cevns makes it complementary to neutrino-electron scattering experiments. Due to its inclusive nature, there is no need to disentangle the sterile
  neutrino mixing from that of the active neutrinos. 
Specifically, we have focused on large liquid argon detectors such as those intended to be installed by the COHERENT Collaboration at the SNS, as well as CCM at the Lujan facility, and the future \cevns program at the ESS.
It is well-known that the admixture of heavy neutrino mediators of neutrino mass generation in the weak charged current induces an effective departure from unitarity
in the lepton mixing matrix.
We have explored how this can affect the initial neutrino fluxes for spallation source experiments, and estimated the projected sensitivities on the unitarity-violating parameters.
In contrast to long-baseline oscillation searches, for the case of short-baseline experiments only the zero-distance effect is relevant.
Our results indicate that future short-baseline \cevns experiments provide a new probe of indirect signatures
  associated to heavy neutrino mediators, with sensitivities competitive with those extracted from global neutrino oscillation data. 
Our main results are shown in Figs.~\ref{fig:chi-diagonal}, \ref{fig:chi2-nondiagonal} and \ref{fig:contour-NU} and summarized in Table~\ref{tab:limits}. 
In long-baseline experiments, the interplay between zero-distance and oscillation effects can make the search for nonunitarity effects more challenging.
A combination of both types of experiments can certainly offer very promising results~\cite{Ge:2016xya}.
All in all, provided the systematic and statistical uncertainties remain under control, the attainable sensitivities to fundamental parameters of the lepton sector
obtained in \cevns experiments will be competitive and complementary to conventional charged-current-based oscillation searches. 

We have also studied the prospects for probing light sterile neutrinos at short-baseline \cevns experiments.
We first verified that the typical baselines of 20--40~m are promising for searches of sterile neutrinos with mass splittings of the order of 1 eV$^2$ (see Figs.~\ref{fig:baseline-mass} and \ref{fig:baseline-mass_future}).
Given the large statistics that can be accumulated by the relevant ton-scale liquid argon detectors, we concluded that CE$\nu$NS-based sterile neutrino searches are feasible,
providing complementary information to the conventional oscillation approaches. The projected exclusion regions in the $(\sin^2 2 \theta_{i4}, \Delta m_{41}^2)$ plane are shown in Fig.~\ref{fig:contour}.
All in all, we have seen that \cevns studies offer a new way to search for light sterile neutrinos, complementary to CC-based short-baseline studies.

\begin{acknowledgments}

This work is supported by the Spanish grants FPA2017-85216-P
(AEI/FEDER, UE), PROMETEO/2018/165 (Generalitat Valenciana) and the
Spanish Red Consolider MultiDark FPA2017-90566-REDC, and by
CONACYT-Mexico under grant A1-S-23238. O.~G.~M has been supported by SNI
(Sistema Nacional de Investigadores). The work of D.~K.~P is cofinanced
by Greece and the European Union (European Social Fund- ESF) through
the Operational Programme <<Human Resources Development, Education and
Lifelong Learning>> in the context of the project ``Reinforcement of
Postdoctoral Researchers - 2nd Cycle" (MIS-5033021), implemented by
the State Scholarships Foundation (IKY). M.~T. acknowledges financial
support from MINECO through the Ram\'{o}n y Cajal contract
RYC-2013-12438.

\end{acknowledgments}

\bibliography{bibliography}
\end{document}